\documentclass[prl,twocolumn,showpacs,preprintnumbers,amsmath,amssymb]{revtex4}
\usepackage{graphicx}% Include figure files
\usepackage{dcolumn}% Align table columns on decimal point
\usepackage{bm}% bold math
\usepackage{psfrag}
\usepackage{epsfig}
\usepackage{amsmath,bbm}
\usepackage{amssymb}
\usepackage{color}
\usepackage{verbatim}
\usepackage{SIunits}
\usepackage{textcomp}
\usepackage{mathbbol}
\usepackage{float}
\usepackage{SIunits}
\newcommand{\ehoch}{\ensuremath{\mathrm{e}^}}

\newcommand{\I}{\ensuremath{\mathrm{i}}}

\newcommand{\bra}[1]{\ensuremath{\langle#1|}}
\newcommand{\ket}[1]{\ensuremath{|#1\rangle}}

\newcommand{\aplus}{\ensuremath{a}^{\dag}}

\newcommand{\om}{\ensuremath{\omega_{\mathrm{m}}}}
\newcommand{\ogate}{\ensuremath{\omega_{\mathrm{G}}}}

\newcommand{\Xgate}{\ensuremath{X_{\mathrm{G}}}}
\newcommand{\Xsave}{\ensuremath{X_{\mathrm{S}}}}
\newcommand{\Hgate}{\ensuremath{H_{\mathrm{G}}}}
\newcommand{\Hsave}{\ensuremath{H_{\mathrm{S}}}}

\newcommand{\gammam}{\ensuremath{\gamma_{\mathrm{m}}}}

\newcommand{\xzpm}{\ensuremath{x_{\mathrm{ZPM}}}}

\newcommand{\bplus}{\ensuremath{b^{\dag}}}

\newcommand{\centerme}[1]{\ensuremath{\begin{matrix}#1\end{matrix}}}

\newcommand{\ggate}{\ensuremath{g_{\mathrm{G}}}}
\newcommand{\gsave}{\ensuremath{g_{\mathrm{S}}}}

\usepackage{xy}
\xyoption{matrix}
\xyoption{frame}
\xyoption{arrow}
\xyoption{arc}

\usepackage{ifpdf}
\ifpdf
\else
\PackageWarningNoLine{Qcircuit}{Qcircuit is loading in Postscript mode.  The Xy-pic options ps and dvips will be loaded.  If you wish to use other Postscript drivers for Xy-pic, you must modify the code in Qcircuit.tex}
\xyoption{ps}
\xyoption{dvips}
\fi

\entrymodifiers={!C\entrybox}

\newcommand{\qw}[1][-1]{\ar @{-} [0,#1]}
\newcommand{\multigate}[2]{*+<1em,.9em>{\hphantom{#2}} \POS [0,0]="i",[0,0].[#1,0]="e",!C *{#2},"e"+UR;"e"+UL **\dir{-};"e"+DL **\dir{-};"e"+DR **\dir{-};"e"+UR **\dir{-},"i" \qw}
\newcommand{\ghost}[1]{*+<1em,.9em>{\hphantom{#1}} \qw}
\newcommand{\Qcircuit}{\xymatrix @*=<0em>}
\begin{document}

\title{Quantum Information Processing with Nanomechanical Qubits}
\author{Simon Rips}
\author{Michael J. Hartmann}
\affiliation{Technische Universit{\"a}t M{\"u}nchen, Physik Department, James Franck Str., 85748 Garching, Germany}

\date{\today}

\begin{abstract}
We introduce an approach to quantum information processing where the information is stored in the motional degrees of freedom of nanomechanical devices. The qubits of our approach are formed by the two lowest energy levels of mechanical resonators which are tuned to be strongly anharmonic by suitable electrostatic fields. Single qubit rotations are conducted by radio frequency voltage pulses that are applied to individual resonators. Two qubit entangling gates in turn are implemented via a coupling of two qubits to a common optical resonance of a high finesse cavity. We find that gate fidelities exceeding 99\% can be achieved for realistic experimental parameters. 
\end{abstract}

% PACS, the Physics and Astronomy Classification Scheme.
\maketitle
%\tableofcontents

% ---------------------------------------------------------------------------
%
Mechanical oscillators are among the most elementary structures that are studied in physics.
Nonetheless they have properties that make them very useful for technological applications.
Their vibrational modes can for example undergo millions of oscillations before the motion is eventually
damped and they can couple to electromagnetic fields
in a very broad frequency range via their polarizability. 
Whereas the latter property has prompted significant effort to build optical to microwave frequency converters \cite{barzanjeh11,regal11},
the enormous Q-factors of mechanical oscillators have been exploited to demonstrate approaches to mechanical computers in the classical domain \cite{Masmanidis,Mahboob}.
The fascinating properties of nanomechanical oscillators furthermore motivated intense research activity towards exploring their quantum regime
which has very recently lead to breakthroughs in cooling such oscillators to their quantum ground states  \cite{OConnell10,Teufel10,Chan11}.

\begin{figure}[h]
\includegraphics[width=\columnwidth]{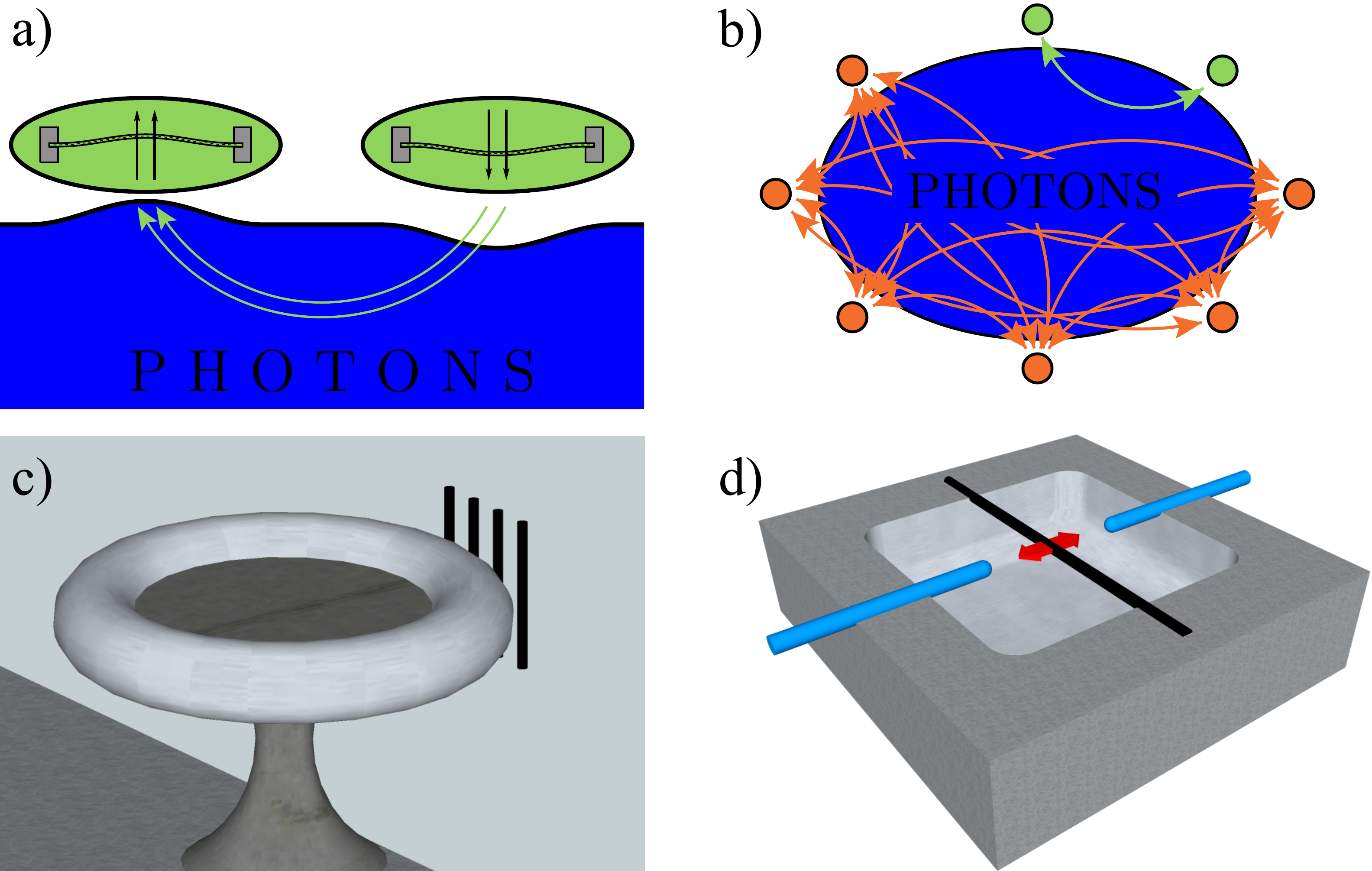}
\caption{Top row: Illustration of nanomechanical qubits interacting via a common photon mode: {\bf a)} The deflection of one resonator locally changes the energy density of the photon field and thus causes a force onto other resonators. {\bf b)} By properly tuning the qubit frequencies, noninteracting subsets can be defined. Bottom row: Possible implementation of the device we envision. {\bf c)} The mechanical vibration of doubly clamped carbon nanotubes couples optomechanically to the evanescent field of a whispering gallery mode in a high finesse micro-toroid. {\bf d)} Electrostatic and radio frequency fields can be applied to each nanotube (black) individually via tip electrodes (blue).}
\label{setup}
\end{figure}

Here we introduce an approach to quantum information processing with mechanical degrees of freedom by making use of the aforementioned
exquisite properties of nanomechanical oscillators. The device we envision, consists of an array of doubly clamped nanobeams that all couple to a common
resonance mode of a high finesse optical cavity, see FIG. \ref{setup} for an illustration and possible setup. Each nanobeam can furthermore be manipulated individually with electrostatic and radio frequency fields that are applied via small tip electrodes. A suitable setup for an implementation are for example carbon nanotubes that couple to the evanescent field of a whispering gallery mode cavity \cite{Steele,Lassagne,Huettel09,Anetsberger09,Anetsberger10,Rips12,Rips12a}, c.f. FIG. \ref{setup}.

% ------------
In our scheme, the constant electric fields applied to individual nanobeams make the mechanical spectrum of each beam anharmonic and allow to tune the respective transition frequencies. For large enough fields, the anharmonicity of the mechanical spectrum becomes comparable to the line-width of the optical cavity resonances. By driving the cavity with a coherent input that is appropriately detuned in frequency from the closest cavity resonance, one can ensure that only one transition between eigenstates of the mechanical motion couples to the cavity. Here, we choose the laser drives such that the cavity couples to the transition between the ground and first excited state of the nanobeam. These two states, denoted $\ket{0}$ and $\ket{1}$, form our nanomechanical qubit.

As compared to previous approaches \cite{Stannigel12,Schmidt12}, our scheme uses the intrinsic nonlinearity of nanomechanical resonators which allows to selectively address the qubit transitions and results in very high gate fidelities.
Local operations respectively single qubit gates are implemented by applying suitable voltage pulses via the tip electrodes.
The optical cavity mode in turn mediates interactions between any pair of mechanical qubits. If such an interaction is active for a suitable time range and combined with pertinent local operations it can implement a fundamental entangling gate, e.g. the so called iSWAP gate \cite{Schuch03}. For many qubits, one can selectively apply such an iSWAP gate to any desired pair of qubits by tuning them to a transition frequency $\omega_{\text{G}}$ while the remaining qubits are tuned to a markedly different transition frequency $\omega_{\text{S}}$ via suitable voltages at the respective tip electrodes. As we show in detail below, at the end of the iSWAP gate on the selected qubits all remaining qubits can be made to return to their initial state and effectively undergo an identity operation. After successfully completing a quantum algorithm, the state of each qubit can be read out by tuning individual qubits to distinct transition frequencies and performing spectroscopy with a weak probe laser, c.f. \cite{Rips12,
Rips12a}.  

With all the above ingredients, the device we envision 
satisfies the requirements for implementing quantum computing \cite{divincenzo00}: 1) It 
has well defined qubits formed by the two lowest energy eigenlevels of strongly anharmonic nanomechanical oscillators and is scalable since multiple nanobeams can couple to one high finesse optical mode. 2) It is initializable in the state $\ket{0,0,\dots,0}$ by cooling the mechanical motion to the ground states, e.g. via side-band cooling \cite{Teufel10,Chan11}. 3) The extremely high Q-factors of nanomechanical oscillators together with a cryogenic environment ensure that the coherence times of the qubits greatly exceed gate operation times. 4) A universal set of gates can be implemented. 5) Spectroscopy on individual qubits together with single qubit rotations allows for high quantum efficiency, qubit-specific readout.

\paragraph{The system}

We consider a system consisting of $N$ nanobeams that are clamped at both ends and thus feature an intrinsic nonlinearity that originates in the stretching of the material associated with its deflection \cite{Carr01,Werner04}. In terms of phonon creation (annihilation) operators $\bplus_j$ ($b_j$), the Hamiltonian of one nanobeam reads,
\begin{equation}
H_{\mathrm{m},j}^{(0)}=\hbar\om\bplus_j b_j+\hbar\frac{\lambda}{2}\left(\bplus_j+b_j\right)^4\, ,
\end{equation}
where $\om$ is the resonance frequency of the harmonic motion and $\lambda = \frac{\beta}{4 \hbar} \xzpm^{4}$ the nonlinear contribution. Here $\beta$ depends exclusively on the dimensions and material properties of the beam whereas $\xzpm$ denotes the amplitude of its zero point motion.
All nanobeams are subject to static and radio frequency electric potentials generated by tip electrodes in the close vicinity of each beam \cite{Wu11,Lee10,Rips12,Rips12a} and couple to a common optical mode of a high finesse cavity. The Hamiltonian of the entire electro-opto-mechanical setup in a frame that rotates at the frequency of the drive laser reads,
% \begin{align}
% \nonumber H=&-\hbar\Delta \aplus a + \hbar g \frac{|\alpha|}{\sqrt{2}} \sum_j X_j + \hbar g X_{\mathrm{c}}\sum_j X_j\\
% &+\sum_j H_{\mathrm{m},j}^{(0)}+\sum_j\left[V^{xy}_j(t)X_j+V^z_j(t)X_j^2\right]\,.
% \label{Hsys}
% \end{align}
\begin{align}
\nonumber H=&-\hbar\Delta \aplus a + \sum_j\hbar g_j \left(\frac{|\alpha|}{\sqrt{2}} + X_{\mathrm{c}}\right) X_j\\
&+\sum_j H_{\mathrm{m},j}^{(0)}+\sum_j\left[V^{xy}_j(t)X_j+V^z_j(t)X_j^2\right]\,.
\label{Hsys}
\end{align}
where $\Delta$ is the detuning between driving laser and cavity resonance and $X_{j} = (b_{j} + \bplus_j)/\sqrt{2}$ the deflection of beam $j$. The photon operators $a$ and $\aplus$ have been shifted by their steady state values, $a\rightarrow \alpha + a$ and a negligible term $\propto \aplus a \sum_{j} X_{j}$ has been dropped. Hence $X_{\mathrm{c}}=(\alpha^{*}a+\alpha\aplus)/\sqrt{2}|\alpha|$ is a photon quadrature with
$\alpha = \frac{\Omega}{2\Delta+\I\kappa}$, where $\Omega$ is the drive amplitude of the laser and $\kappa$ the photon decay rate of the cavity.

The potentials $V_{j}^{xy,z}(t) = V_{j,0}^{xy,z} + V_{j,1}^{xy,z}(t)$ describe constant, $V_{j,0}^{xy,z}$, and time dependent, $V_{j,1}^{xy,z}(t)$, gradient forces caused by the voltages applied to the tip electrodes. The constant parts tune the equilibrium positions of the mechanical oscillators via $V_{j,0}^{xy}$ and their spectrum via $V_{j,0}^{z}$, whereas the time dependent parts $V_{j,1}^{xy}(t)$ can implement single qubit rotations.
We always choose $V_{j,0}^{xy} = -\hbar g_j \frac{|\alpha|}{\sqrt{2}}$ such that the nanobeams remain undeflected.
$g_j = 2|\alpha|x_{\mathrm{zpm},j} G_0$ is the optomechanical coupling that can be controlled by the amplitude of the laser drive,
where $G_0=\partial\omega/\partial X$ is the cavity's frequency shift per resonator deflection.
We use the radio frequency fields $V_{j,1}(t)$ and the couplings $g_j$ as controls to perform gate operations.

In a realistic experimental situation, the mechanical motion will be subject to damping at a rate $\gammam$ and cavity photons will be lost at a rate $\kappa$. The full dynamics of our system that takes these incoherent processes into account can thus be described by the master equation,
\begin{align}
\label{mg}
 \dot{\varrho} = & -i \left[H, \varrho \right] \\
 & + \frac{\gammam}{2} \sum_{j} \left[\overline{n} \, \mathcal{D}_{\uparrow,j}(b_{j}) + (\overline{n} + 1) \mathcal{D}_{\downarrow,j}(b_{j}) \right] + \frac{\kappa}{2} \mathcal{D}_{\downarrow,c}(a)  \nonumber\,,
 \end{align}
where $\overline{n}$ is the thermal phonon number at the environment temperature $T$ and the dissipators read $\mathcal{D}_{\downarrow}(y) = 2 y \varrho y^{\dagger} - y^{\dagger} y \varrho - \varrho y^{\dagger} y$
and $\mathcal{D}_{\uparrow}(y) = 2 y^{\dagger} \varrho y - y y^{\dagger} \varrho - \varrho y y^{\dagger}$.

\paragraph{Nanomechanical qubits}
The qubits we consider are formed by the two lowest energy levels of our nanomechanical beams. Mechanical resonators feature a small instrinsic anharmonicity in any deflection mode that can be enhanced by electrostatic gradient forces, see \cite{Rips12,Rips12a} for details. Suitably tuned electrostatic fields generate an inverted harmonic potential of the form $V \propto - X^{2}$ that counteracts the harmonic part of the elastic restoring force. The electrostatic potential thus reduces the stiffness respectively softens the mechanical resonance mode and hence reduces its frequency.
Since the deflection per phonon, $\xzpm$, is proportional to $\om^{-1/2}$ a reduction of $\om$ causes a significant enhancement of the nonlinearity as $\lambda \propto \xzpm^{4} \propto \om^{-2}$, c.f. \cite{Rips12,Rips12a}.

\begin{figure}
\includegraphics[width=\columnwidth]{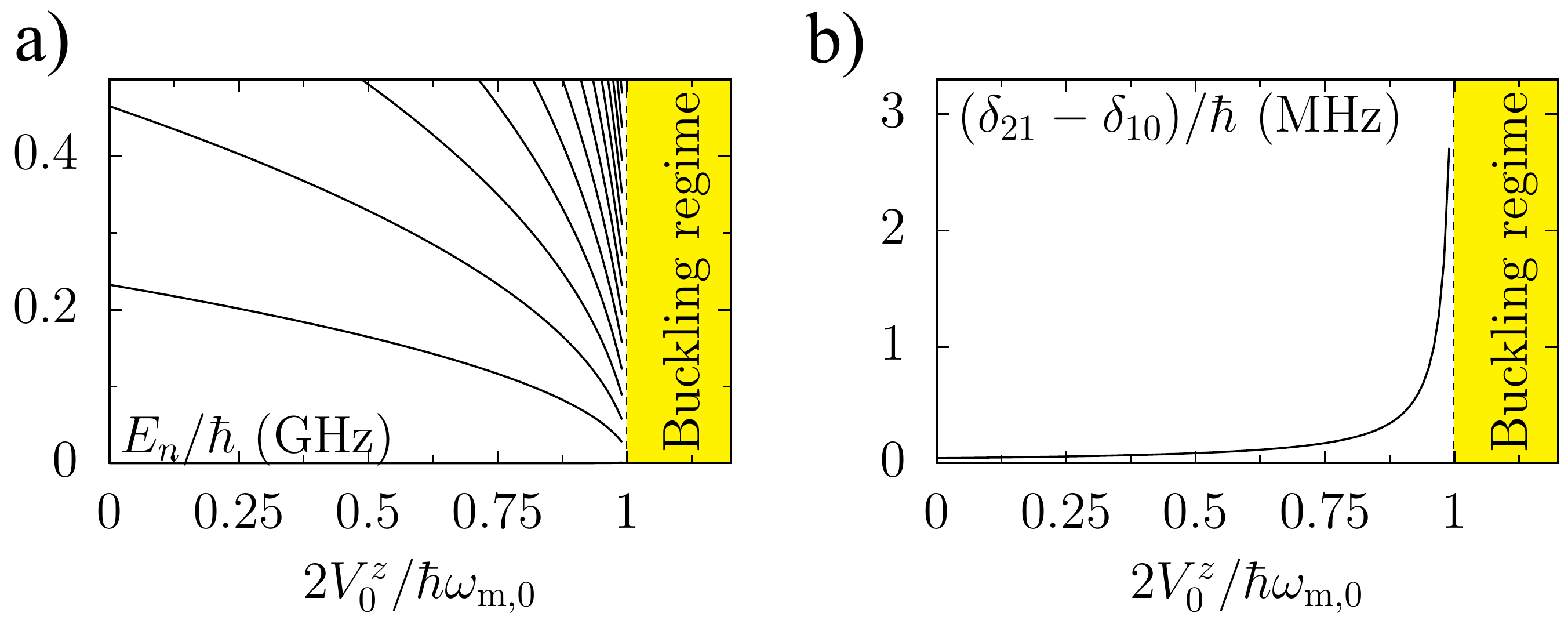}
 \caption{{\bf a)} Nonlinear phonon spectrum of a nanobeam in the presence of a softening field $V_{0}^{z}$. For $2 V_{0}^{z} > \hbar \omega_{m,0}$, the beam enters the bistable buckling regime.  {\bf b)} Effective nonlinearity of the phonon spectrum
 $\delta_{21}-\delta_{10}$. A sufficiently high value of $\delta_{21}-\delta_{10}$ suppresses unwanted processes of the type $\ket{11...}\rightarrow\ket{20...}$.}
 \label{spectrum}
\end{figure}

It is useful to consider a ``tuned'' mechanical Hamiltonian, $H_{\mathrm{m},j} = H_{\mathrm{m},j}^{(0)} + \sum_j V^{z}_{j,0} X_j^{2}$, that includes the constant
parts of the $V^{z}_{j}$. We write $H_{\mathrm{m},j}$ and the deflection $X_{j}$ of each mechanical resonator in the eigenbasis of $H_{\mathrm{m},j}$,
%\begin{equation}
%H_{\mathrm{m},j}=\sum_nE_n\ket{n}_j\bra{n}_{j}\,, \quad  X_j=\sum_{nm}X_{nm}\ket{n}_j\bra{m}_j\,.
%% H_{\mathrm{m},j}=\frac{P_{j}^{2}}{2 m} + \frac{m}{2} \om^{2} X_{j}^{2} + \hbar\lambda X_{j}^4=\sum_nE_n\ket{n}_j\bra{n}_{j}\,,
%\label{MechHamilton}
%\end{equation}
i.e. $H_{\mathrm{m},j}=\sum_nE_{n,j}\ket{n}_j\bra{n}_{j}$ and $X_j=\sum_{nm}X_{nm,j}\ket{n}_j\bra{m}_j$, 
where $H_{\mathrm{m},j}\ket{n}_{j} = E_{n,j} \ket{n}_{j}$. The $E_{n,j}$ and $X_{nm,j}$ can be found numerically, see FIG. \ref{spectrum}a.

Provided the nonlinearity per phonon is large enough and the mechanical motion is cooled to the groundstate, we can restrict our analysis to the two lowest energy levels $\ket{0}_{j}$ and $\ket{1}_{j}$ of each resonator which form our qubits.
The nonlinearity of the qubits, i.e. the fact that the qubit transition energies $\delta_{10,j}=E_{1,j}-E_{0,j}$ 
differ significantly from transition energies between higher states $\delta_{nm,j}=E_{n,j}-E_{m,j}$, especially from $\delta_{21,j}$ (see FIG. \ref{spectrum}b),
ensures that the restriction to these two levels remains a very accurate approximation throughout the gate sequences.
To explain the gate operations in detail we now switch to an interaction picture with respect to 
$H_0=-\hbar\Delta\aplus a+\sum_jH_{\mathrm{m},j}$. 

\paragraph{Local Operations and Single Qubit Gates}

Local operations, that is rotations about the $\sigma_x$-, $\sigma_y$- and $\sigma_z$-axes are realized by applying time dependent gradient forces as encoded in the potentials $V^{xy}_j(t)$ and $V^z_j(t)$ in eq. \eqref{Hsys}.

$\sigma_z$-rotations are obtained by temporarily shifting the qubit transition frequency $\delta_{10}$ which makes the qubit rotate at a different rate and hence collect a phase shift. This is achieved by tuning the softening fields that control the mechanical frequency to a different value for a suitable time which
adds a term $V^{z} X^2$ to the Hamiltonian that shifts the qubit frequency $\delta_{10}\rightarrow\delta_{10} + \delta_{10}^{(1)}(t)$
(We skip the qubit index $j$ throughout the discussion of local operations).
For $\int\delta_{10}^{(1)}{\rm d}t = \phi/2$, this procedure implements the operation
\begin{equation}
 \ehoch{-\I\int V^{z}(t) X^2{\rm d}t/\hbar}\approx\ehoch{-\I\sigma_z\phi/2}\equiv\centerme{\Qcircuit@C=.5em @R=.7em{&\qw&&[\,\phi\,]_z&&\qw}}\,.
\end{equation}

$\sigma_x$- and $\sigma_y$-rotations can be implemented by applying gradient forces related to $V^{xy}$.
If the potential $V^{xy}(t)$ is a pulse that is modulated by an oscillation at the qubit frequency $V^{xy}(t)=\cos(\delta_{10}t+\theta)\tilde V^{xy}$, one finds 
$\tilde V^{xy}\cos(\delta_{10}t+\theta)\ehoch{\I H_0 t/\hbar}X\ehoch{-\I H_0 t/\hbar}\approx\tilde V^{xy} \times$ $\times \frac{X_{01}}{2}\left[\cos(\theta)\left(\sigma^{01}+\sigma^{10}\right)+\I\sin(\theta)\left(\sigma^{01}-\sigma^{10}\right)\right]$,
where $\sigma^{ab} = \ket{a}\bra{b}$ and we have dropped off-resonant contributions. Thus, for $\theta=0$ and $\int{\rm d}t\tilde V^{xy}(t)=\phi/X_{01}$ one finds
\begin{equation}
\ehoch{-\I\int{\rm d}t V^{xy}(t)X}\approx\ehoch{-\I\sigma_x\phi/2}\equiv\centerme{\Qcircuit@C=.5em @R=.7em{&\qw&&[\,\phi\,]_x&&\qw}}\,,
\end{equation}
and for $\theta=\pi/2$  and $\int{\rm d}t\tilde V_{xy}(t)=\phi/X_{01}$
\begin{equation}
\ehoch{-\I\int{\rm d}t V^{xy}(t)X}\approx\ehoch{-\I\sigma_y\phi/2}\equiv \centerme{\Qcircuit@C=.5em @R=.7em{&\qw&&[\,\phi\,]_y&&\qw}}\,.
\end{equation}

\paragraph{Effective interactions and two-qubit gates}
In order to implement entangling two-qubit operations we employ an indirect interaction between mechanical resonators that is mediated by an optical cavity mode \cite{Hartmann08}. Here, the laser is sufficiently far detuned from any resonance, such that 
$g_j\ll|\Delta\pm\delta_{nm,j}|$.
For an initial state in the qubit subspace formed by $\ket{0}_{j}$ and $\ket{1}_{j}$, this condition ensures that the dynamics is restricted to this subspace.
To perform a gate operation on two qubits, e.g. with indices $j=\{1,2\}$ (``gate qubits''), we tune those two qubits to a common transition frequency $\omega_{\mathrm{G}}$ and hence common coupling $g_{\text{G}}$, while tuning the remaining qubits (``saved qubits'') to a sufficiently different transition frequency $\omega_{\mathrm{S}}$ and coupling $g_{\text{S}}$ such that $|\omega_{\mathrm{G}}-\omega_{\mathrm{S}}|\gg g_{\text{G}}, g_{\text{S}}$ and interactions between gate qubits and saved qubits are strongly suppressed. 
To explain the working principle of this gate, we consider a scenario where $V^{xy,z}_{j,1} = 0$.
An adiabatic eliminiation of the photons together with a rotating wave approximation yields the effective Hamiltonian
% \begin{equation}
%  H_{\mathrm{eff}}\approx H_{\mathrm{G}}+H_{\mathrm{S}}\,.
%  \label{Heff}
% \end{equation}
% with 
% $
% H_{\mathrm{G}}=\hbar \frac{g_{\mathrm{G}}^2 \Delta X_{\mathrm{G}}^2}{\Delta^2-\omega_{\mathrm{G}}^2}\left(\sigma^{01}_1 \, \sigma^{10}_2 + \text{H.c.} \right)+\sum_{i=1}^2 H_{\text{G},i}
% $
% and
% $
% H_{\mathrm{S}}=\hbar \frac{g_{\mathrm{S}}^2 \Delta X_{\mathrm{S}}^2}{\Delta^2-\omega_{\mathrm{S}}^2}\sum_{i\neq j> 2}\left(\sigma^{01}_{i} \, \sigma^{10}_{j} + \text{H.c.} \right)
% +\sum_{i>2}H_{\text{S},i}
% $,
\begin{align}
 \label{Heff}
H_{\mathrm{eff}}&\approx H_{\mathrm{G}}+H_{\mathrm{S}}\\
H_{\mathrm{G}}&=\hbar \frac{g_{\mathrm{G}}^2 \Delta X_{\mathrm{G}}^2}{\Delta^2-\omega_{\mathrm{G}}^2}\left(\sigma^{01}_1 \, \sigma^{10}_2 + \text{H.c.} \right)+\sum_{j=1}^2 H_{\text{G},j} \nonumber \\
H_{\mathrm{S}}&=\hbar \frac{g_{\mathrm{S}}^2 \Delta X_{\mathrm{S}}^2}{\Delta^2-\omega_{\mathrm{S}}^2}\sum_{i\neq j> 2}\left(\sigma^{01}_{i} \, \sigma^{10}_{j} + \text{H.c.} \right)
+\sum_{j>2}H_{\text{S},j}, \nonumber
 %\label{Heff2}
\end{align}
where 
$H_{\text{G/S},j} = \hbar \frac{g_j^2}{2}\left(c^0_{\text{G/S}}\sigma^{00}_{j}+c^1_{\text{G/S}}\sigma^{11}_{j}\right)$ and $\sigma^{ab}_{j} = \ket{a}\bra{b}_j$.
Here, $X_{\mathrm{G/S}}$ denotes the displacement matrix element $X_{01,j}$ and $c^n_{\mathrm{G/S}}=\sum_m\tfrac{X_{nm,j}^2}{\Delta+\delta_{nm,j}}$ the interaction induced energy shifts for the gate qubits ($j=1,2$) and saved qubits ($j>2$) respectively, see supplementary material \cite{supplement}. The corresponding phase shifts can be reversed after the gate operation by appending a suitable $\sigma_z$-rotation.

The time evolution of the Hamiltonian \eqref{Heff} can be used to perform an iSWAP operation on distinct gate qubits, while the other qubits are unaffected. This can be achieved by decomposing the iSWAP operation into $\sqrt{\pm\text{iSWAP}}$ operations, using the identity
\begin{align}
\hspace{-0.08cm}
\centerme{
\Qcircuit @C=0.5em @R=.5em {
& \multigate{1}{\sqrt\mathrm{iSWAP}} &\qw&*-<1.0em>{[\pi]_z}&\qw&\multigate{1}{\sqrt\text{-iSWAP}}&\qw&*-<1.0em>{[\text{-}\pi]_z}&\qw
\\
& \ghost{\sqrt\mathrm{iSWAP}} 		&\qw &\qw &\qw&\ghost{\sqrt\text{-iSWAP}}&\qw&\qw&\qw
}}&=\centerme{\Qcircuit @C=0.5em @R=.5em {
& \multigate{1}{\mathrm{iSWAP}}	&\qw
\\
& \ghost{\mathrm{iSWAP}} 	&\qw	
}}
\label{SqrtiSWAP}
\end{align}
Here, the $\sqrt{\pm\text{iSWAP}}$ operations are implemented by choosing laser drive pulses such that
$\frac{\Delta X_{\mathrm{G}}^2}{\Delta^2-\omega_{\mathrm{G}}^2} \int g_{\mathrm{G}}(t)^2 {\rm d}t = \pm\frac{\pi}{4}$ respectively.
The $\sqrt{\text{iSWAP}}$ and the $\sqrt{-\text{iSWAP}}$ operations thus only differ by the sign of the employed detuning $\Delta$. During the $\sqrt{\text{iSWAP}}$ operation on the gate qubits, the saved qubits are subject to the dynamics generated by $\Hsave$ (compare FIG. \ref{setup}b). Yet since there are no local $\centerme{\Qcircuit@C=.5em @R=.7em{&\qw&&[\,\pi \,]_z &&\qw}}$ operations on the saved qubits, this evolution is reversed during the subsequent $\sqrt{-\text{iSWAP}}$ operation on the gate qubits and the saved qubits return back to their initial states.

\begin{figure}
\includegraphics[width=\columnwidth]{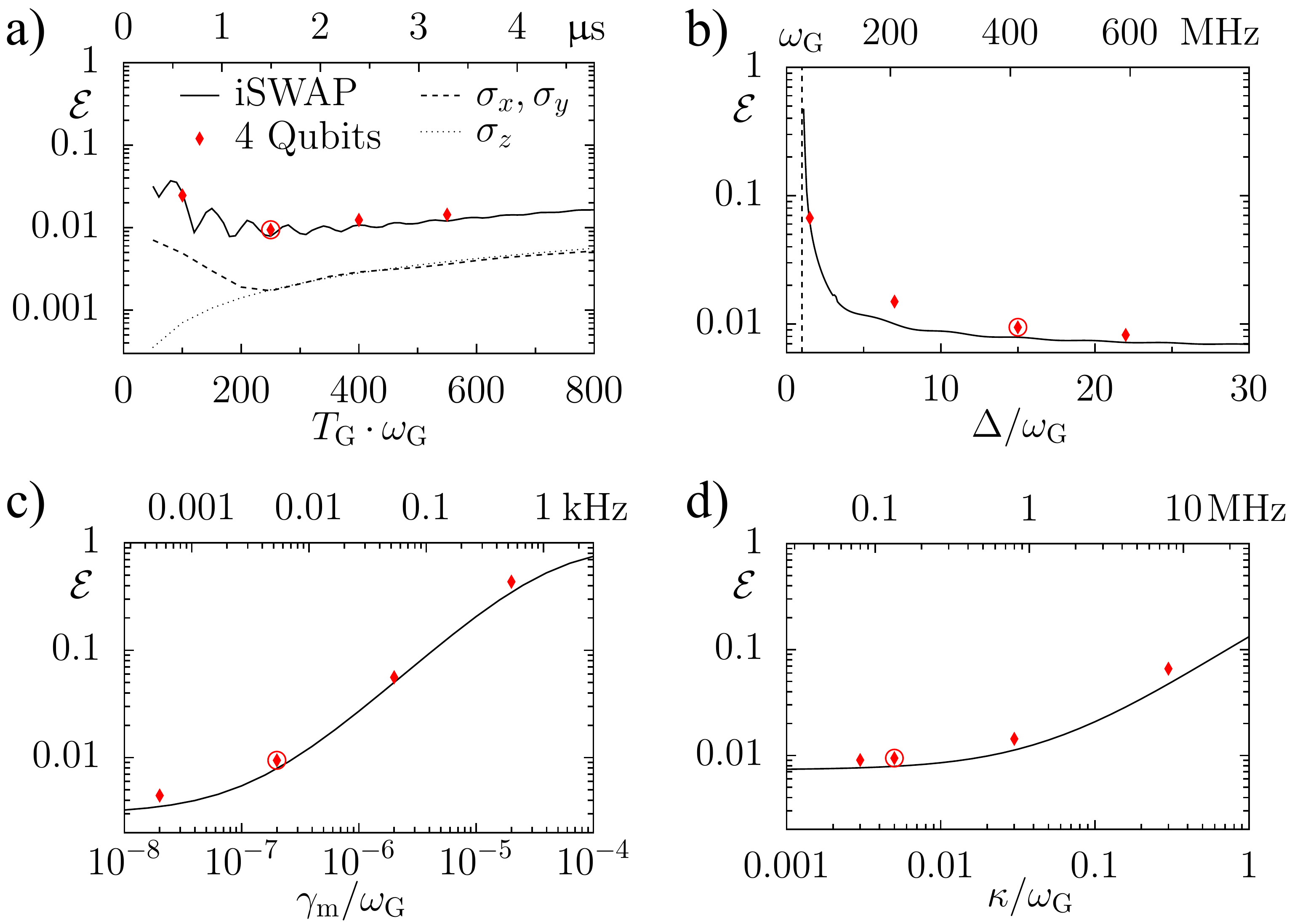}
\caption{Error $\mathcal{E}=1-F$ of gate operations. Except for the quantities on the horizontal axis all parameters are as in the setting described in the main text. Solid lines show results for two and red dots for four qubits. Highlighted red dots correspond to the parameter example discussed in the text. {\bf a)} Error as a function of the gate time for local and an entangling operations. Note that changing the gate time also changes the required coupling rate. {\bf b)} Error as a function of the laser detuning. {\bf c)} and {\bf d)} illustrate the influence of damping for the photons and the resonators, repectively.}
\label{results}
\end{figure}

\paragraph{Numerical results}
We analyzed the fidelites of the quantum gates we propose by numerically solving (\ref{mg}) involving two and four qubits where we included the lowest three levels $\ket{0},\ket{1},\ket{2}$ for each resonator respectively qubit. To find an estimate for the fidelity of the gate operations we compute the fidelity $f(\sigma, \rho) = \text{Tr}(\sqrt{\sqrt{\rho}\,\sigma\sqrt{\rho}})$ of the desired target state $\sigma$ with the actual state $\rho$ that results from the full dynamical evolution for several initial states. In order to confirm that relative phases of the states evolve as desired, we use all states of the form $|\phi\rangle = (|j,k,\dots\rangle + |0,0,\dots\rangle)/\sqrt{2}$ ($j,k,\ldots = 0,1$) as initial states and average the fidelity $f$ over these,
$F = \overline{f(\sigma, \rho)}\big|_{\text{all} \, \ket{\phi}}$.
FIG. \ref{results} shows the resulting gate error $\mathcal{E}=1-F$ as a function of the gate time $T_{\text{G}}$, the laser detuning $\Delta$ and the mechanical and optical dissipation rates $\gamma_{\text{m}}$ and $\kappa$. To demonstrate the scalability of our approach we present results for two qubits (solid lines) as well as for four qubits (red dots). For the four qubit case, the gate is applied to two qubits while the remaining two return to their initial state. These results cleary show the excellent scaling properties of our approach.

The result given in FIG. \ref{results} are found for ($10,0$)-carbon nanotubes of length $L=\unit{300}{\nano\meter}$ and radius $R=\unit{0.39}{\nano\meter}$, coupled to the evanescent field of a micro toroid cavity, see \cite{Rips12,Rips12a} for details. By softening the mechanical resonance frequency of the gate qubits down to $\ogate/2\pi=\unit{26.6}{\mega\hertz}$,
we achieve a nonlinearity of $(\delta_{21}-\delta_{10})/2\pi=2.71\,$MHz. The optomechanical coupling of carbon nanotubes can be dramatically enhanced by employing a cavity resonance in the vicinity of recently demonstrated excitonic resonances \cite{hoegele,Wilson-Rae09,supplement}. We choose a cavity frequency that is far enough from such a resonance to sufficiently suppress additional absorption and estimate that an optomechanical coupling rate of $g_{\text{G}}/2\pi=8.73\,$MHz can be achieved with a laser drive of $\unit{5.30}{\watt}$ input power that is detuned by $\Delta/2\pi=\unit{0.399}{\giga\hertz}$ from the cavity frequency. For a gate time of $T_{\text{G}}=\unit{1.49}{\micro\second}$ such coupling suffices for the $\sqrt{\mathrm{iSWAP}}$ operation asuming a rectangular pulse. Furthermore we asume an ambient temperature of $\unit{20}{\milli\kelvin}$, a mechanical $Q$-factor of $5\cdot10^{6}$ for the tuned resonator \cite{Huettel09} and a total photon decay rate (including intrinsic cavity losses and 
losses 
induced by the exciton resonance) of $\kappa/2 \pi=\unit{133}{\kilo\hertz}$  \cite{Kippenberg04,Poellinger09}. Note that for our conditions with $\kappa^{2}/\Delta^{2} = 8 \times 10^{-8}$ this corresponds to an absorbed power of $\sim\unit{18}{\nano\watt}$ that is compfortly compatible with a cryogenic environment \cite{Riviere11}. Moreover even higher finesses and lower absorption could be attained in crystalline resonators \cite{Hofer10}.
For tuning the mechanical resonances, electrostatic fields of $\unit{76}{\volt\per\micro\meter}$ are required, whereas for the local operations $\sigma_{x,y}[\phi]$, we find radio frequency field amplitudes of $\sim \unit{1}{\volt\per\micro\meter}$, depending on operation time and angle $\phi$. For the static fields that compensate the photon induced shift of the equilibrium position during an entangling operation, $\unit{6.1}{\volt\per\micro\meter}$ are needed and the $\sigma_z[\phi]$ operations require a gate time  $>\phi/\ogate$ to avoid buckling the resonator. 
Importantly, decoherence due to electric noise in the electrodes, such as Johnson-Nyquist or $1/f$ noise, is in our setup negligible compared to the mechanical damping $\gammam \overline{n}$ \cite{supplement}.  

\paragraph{Conclusions}
In summary, we have introduced an approach to quantum information processing with nanomechanical qubits.
Our approach is realizable with current or near future experimental technology. 
Importantly, the nanomechanical resonators in our approach show very promising scalability properties and are not vulnerable to fluctuations of background charges such as superconducting qubits. The performance of our scheme could be improved further with optimized laser and radio frequency control pulses. Alternative platforms for an implementation could consist of stiff nanobeams with high optomechanical couplings such as
photonic crystal nanobeam cavities in diamond \cite{Hausmann12}. For building even larger scale devices our scheme could also be integrated into optomechanical networks of multiple cavities \cite{Stannigel10}.

\paragraph{Acknowledgements}
The authors thank Ignacio Wilson-Rae and Alexander H\"ogele for fruitful discussions.
This work is part of the Emmy Noether project HA 5593/1-1 and the CRC 631, both funded by the German Research Foundation, DFG.

\widetext

\appendix

\section{Supplementary Material}

\subsection{Derivation $H_{\mathrm{eff}}$}

The effective Hamiltonian given in equation (7) of the main text describes the dynamics of the qubit array while performing entangling gate operations. It is obtained for large detunings $\Delta$, where the cavity can be adiabatically eliminated. Here we provide a derivation of the Hamiltonian (7) that describes the dynamics at timescales larger than $\Delta^{-1}$.
The Schr\"odinger picture Hamiltonian of our setup reads 
\begin{equation}
 H=-\Delta \aplus a + \sum\limits_i g_i(t) X_{\mathrm{c}}X_i+\sum\limits_iH_{\mathrm{m},i}\,.
 \end{equation}
In changing to the interaction-picture with respect to $-\Delta\aplus a+\sum_iH_{\mathrm{m},i}$, we have to account for the mechanical nonlinearity and thus employ the decomposition of $X_i$ in energy eigenstates  
 \begin{equation}
 H_I(t)=\sum\limits_{nm,i}\frac{g_i(t)}{\sqrt{2}}\left(\aplus\ehoch{-\I\Delta t}+a\ehoch{\I\Delta t}\right)X_{nm,i}\ehoch{\I\delta_{nm,i}t}\ket{n}_{i}\bra{m}_i
 \label{H_I}
\end{equation}
In a regime where $g_i\ll|\Delta\pm\delta_{nm,i}|$, the effective Hamiltonian is given by \cite{James},
\begin{equation}
 H_{\mathrm{eff}}=\frac{1}{\I}H_I(t)\int^tH_I(t'){\rm d}t'
\end{equation}
For a large detuning $\Delta$, the optomechanical interaction can not efficiently populate the cavity. We thus assume an empty cavity mode and 
drop fast rotating terms ${\aplus}^2$ and $a^2$, 
%all 
which leads to
\begin{align}
\nonumber H_{\mathrm{eff}}=&\sum\limits_{\substack{nmlk\\i\neq j}}\frac{g_ig_jX_{nm,i}X_{lk,j}}{2(\Delta-\delta_{lk,j})}\ehoch{\I(\delta_{nm,i}+\delta_{lk,j})t}\ket{n}_{i}\bra{m}_i \, \ket{l}_{j}\bra{k}_j\\
  &+\sum\limits_{\substack{nmk\\i}}\frac{g_i^2X_{nm,i}X_{mk,i}}{2(\Delta-\delta_{mk,i})}\ehoch{\I\delta_{nk,i}t}\ket{n}_{i}\bra{k}_i \, ,
\end{align}
where indices $i$ and $j$ label the resonators (qubits) and indices $n,m,l$ and $k$ label internal resonator levels.
Now we account for the different transition frequencies of ``gate qubits'' and ``saved qubits'' as described in the main text and drop all off-resonant contributions, since $g_i^2/\Delta\ll\left\{\delta_{nm,i},\Delta\right\}$. This separates $H_{\mathrm{eff}}$ into two noninteracting parts $\Hgate$ and $\Hsave$, acting separately on gate and saved qubits. Since $\delta_{1n,i}\neq\delta_{01,j}$ for any qubits $i,j$ and for $n>1$, we can furthermore separate the terms in $H_{\mathrm{eff}}$ that act the qubit subspace from the remainder.
Denoting $g_{i} = g_{\text{G}}$, $X_{01,i} = X_{\text{G}}$ and $\delta_{10,i} = - \delta_{01,i} =\omega_{\text{G}}$ for $i= 1,2$ as well as $g_{i} = g_{\text{S}}$, $X_{01,i} = X_{\text{S}}$ and $\delta_{10,i} = - \delta_{01,i} = \omega_{\text{S}}$ for $i>2$ we arrive at, 
\begin{align}\label{eq:Heff}
 \Hgate=&\frac{1}{2}\left(\frac{\ggate^2\Xgate^2}{\Delta + \omega_{\text{G}}}+\frac{\ggate^2\Xgate^2}{\Delta-\omega_{\text{G}}}\right)\left\{\ket{1}_{1}\bra{0}_1 \, \ket{0}_{2}\bra{1}_2+\text{H.c.}\right\}\\
 &+\frac{1}{2}\sum\limits_{\substack{m\\i=1,2}}\left(\frac{\ggate^2X_{0m,i}^2}{\Delta+\delta_{0m,i}}\ket{0}_{i}\bra{0}_i+\frac{\ggate^2X_{1m,i}^2}{\Delta+\delta_{1m,i}}\ket{1}_{i}\bra{1}_i\right) \nonumber \\
  \Hsave=&\frac{1}{2}\left(\frac{\gsave^2\Xsave^2}{\Delta+\omega_{\text{S}}}+\frac{\gsave^2\Xsave^2}{\Delta-\omega_{\text{S}}}\right)\sum\limits_{i\neq j>2}\left\{\ket{1}_{i}\bra{0}_i \, \ket{0}_{j}\bra{1}_j+\text{H.c.}\right\} \nonumber \\
 &+\frac{1}{2}\sum\limits_{\substack{m\\i>2}}\left(\frac{\gsave^2X_{0m,i}^2}{\Delta+\delta_{0m,i}}\ket{0}_{i}\bra{0}_i+\frac{\gsave^2X_{1m,i}^2}{\Delta+\delta_{1m,i}}\ket{1}_{i}\bra{1}_i\right) \nonumber 
\end{align}
Note that the first line in $\Hgate/\Hsave$ represents swapping of qubits, while the second line represents an interaction induced shift of the transition frequency. 
Equation (7) of the main text follows directly form equation (\ref{eq:Heff}).

\subsection{CNT-excitons and polarizability}

\begin{figure}[h]
 \centering\includegraphics[width=0.4\textwidth]{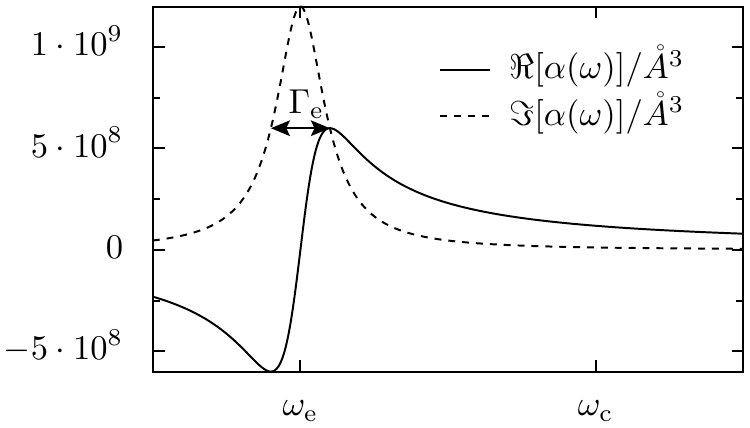}
  \caption{Sketch of the real and imaginary part of the dynamical polarizability at an excitonic resonance. We choose a cavity frequency $\omega_{\text{c}}$ that is sufficiently detuned from the resonance to suppress losses induced by $\Im(\alpha)$.}
 \label{exzitons}
\end{figure}

The optomechanical coupling of the nanotube to the cavity is mediated by gradient forces, originating from the evanescent cavity field and acting on the nanoresonator. Therefore, the strength of optomechanical coupling is proportional to the polarizability $\alpha$ of the nanotube. Here, we make use of the fact, that at the scale of $\unit{}{\micro\meter}$ optical wavelength, the polarizability can be significantly enhanced due to the presence of excitonic resonances \cite{hoegele}. Near an excitonic resonance of frequency $\omega_{\mathrm{e}}$ and linewidth $\Gamma_{\mathrm{e}}$, the polarizability can be estimated by
\begin{equation}
 \alpha(\omega)=\frac{\mathrm{e^2f}}{m_{\mathrm{e}}}\frac{1}{\omega_{\mathrm{e}}^2-\omega^2+\I\omega\Gamma_{\mathrm{e}}}\,,
\end{equation}
where $f\approx10-100$ is a typical oscillator strength in these systems \cite{hoegele_private}. With $\hbar\omega_{\mathrm{e}}\approx \unit{1.4}{\electronvolt}$ and $1/\Gamma_{\mathrm{e}}\approx\unit{1}{\nano\second}$ \cite{hoegele}, this leads to a potential amplification of the polarizability and hence, optomechanical coupling strength, by a factor $>10^6$ (at the maximum of $\Re(\alpha)$) as compared to the static value of $\unit{10-100}{\angstrom^3}$. However, to minimize additional photon losses introduced by the imaginary part $\Im(\alpha)$, it is preferable to choose the cavity to be detuned from the exciton resonance, compare FIG. \ref{exzitons}. We choose a detuning of $\omega_{\mathrm{c}}-\omega_{\mathrm{e}}=10^{-4}\omega_{\mathrm{e}}$, which yields an enhancement factor of roughly $2\times10^4$, and additional losses of $\kappa_{\mathrm{e}}\approx\unit{60}{\kilo\hertz}$. These losses add to the intrinsic cavity losses to yield a broadened cavity linewidth $\kappa$, which we considered in our 
calculatios.

\subsection{Electric noise}

We here give an estimate of decoherence rates for the qubits induced by noise in the electric gradient fields. Such noise might originate from voltage flucuations $\delta U$ due to the electrodes resistance (Johnson-Nyquist noise) or from moving charges on the chip surface ($1/f$-noise). We calculate the respective decoherence rates $\Gamma_{\delta U}$ and $\Gamma_{1/f}$ from the corresponding noise spectra by 
\begin{equation}
 \Gamma_i=\frac{\xzpm^2}{\hbar^2}S_{\delta F_i}(\omega_{\mathrm{m}}) \quad \text{with} \quad S_{\delta F_i}(\omega)=\text{Re}\int\limits_0^{\infty}{\rm d}\tau\left\langle \delta F_i(\tau)\delta F_i(0)+\delta F_i(0)\delta F_i(\tau)\right\rangle\ehoch{\I\omega\tau}\,,
\end{equation}
where $\delta F_i$ is the force fluctuation acting on the resonator. The electric field gradient force acting on a resonator can be expressed by
\begin{equation}
 F_{\mathrm{el.}}= \frac{\alpha}{2} \, \frac{\partial}{\partial X} \int E^{2} {\rm d}l \approx\alpha a E\left(\frac{E}{a}\right)\,,
\end{equation}
where we estimated the field gradient at a distance $a$ from the electrode by $E/a$ and used the fact that the field mainly acts on the nanotube in a region of length $a\ll L$. For a fluctuating field $E+\delta E$, the force fluctuations are then given by
\begin{equation}
\delta F_i = 2\alpha E \delta E\,.
\end{equation}
Thus, the resulting decoherence rates read
\begin{equation}
 \Gamma_i\approx\frac{\xzpm^2}{\hbar^2}S_{\delta F_i}=4\frac{\xzpm^2}{\hbar^2}\alpha^2E^2S_{\delta E}\,,
\end{equation}
where $S_{\delta E}$ is the noise spectrum for the electric field fluctuations.

\paragraph{Johnson-Nyquist noise}

For Johnson-Nyquist noise \cite{Nyquist}, we have fluctuating voltages $\delta U$ with
\begin{equation}
S_{\delta U}\simeq4k_{\mathrm{B}}TR \quad \text{and thus} \quad S_{\delta E}\approx S_{\delta U}/a^2\,,
\end{equation}
for an ambient temperature $T$ and an electrode resistance $R$.
For our setup we thus find $\Gamma_{\delta U}/R < \unit{10^{-2}}{\hertz\per\ohm}$, which is well below the mechanical damping $\gammam \overline{n}\approx\unit{507}{\hertz}$ for a large range of possible resistances.

\paragraph{$1/f$-noise}

The origin of $1/f$-noise is usually associated with surface charge fluctuations in the device. An electric field noise density $S_{E}(\omega/2\pi=\unit{3.9}{\kilo\hertz})\approx\unit{4}{\square\volt\rpsquare\meter\reciprocal\hertz}$ has been measured at $T=\unit{300}{\kelvin}$ and at a distance of $\unit{20}{\nano\meter}$ between a charged resonator and a gold surface \cite{Stipe}. For a scaling $S_{E}(\omega)\sim T/\omega$  \cite{Stipe,Rabl} this corresponds to $S_{E}\approx\unit{5\cdot10^{-8}}{\square\volt\rpsquare\meter\reciprocal\hertz}$ for our conditions with $T=\unit{20}{\milli\kelvin}$ and $\omega_{\mathrm{m}}/2\pi=\unit{26.6}{\mega\hertz}$. For the associated decoherence rate we thus find $\Gamma_{1/f}<\unit{0.1}{\hertz}$, which is again well below the mechanical damping $\gammam \overline{n}$.

This results are also corroborated by recent estimates that were obtained for a related setup \cite{Wilson-Rae09}.

\paragraph{Acknowledgements}
The authors thank Ignacio Wilson-Rae and Alexander H\"ogele for fruitful discussions.
This work is part of the Emmy Noether project HA 5593/1-1 and the CRC 631, both funded by the German Research Foundation, DFG.

\end{document}